\documentclass[twocolumn,superscriptaddress,pra]{revtex4}
\usepackage{amsmath, amsthm, amssymb}
\usepackage{graphicx}
\usepackage{dcolumn}
\usepackage{bm}
\usepackage{color} 

\newcommand{\ket}[1]{| #1 \rangle}
\newcommand{\bra}[1]{\langle #1 |}

\newtheorem{theorem}{Theorem}

\begin{document}

\title{Fully nonlocal, monogamous, and random genuinely multipartite quantum correlations}

\author{Leandro Aolita}
\affiliation{ICFO-Institut de Ci\`{e}ncies Fot\`{o}niques, Av. Carl Friedrich Gauss 3, E-08860 Castelldefels (Barcelona), Spain}
\author{Rodrigo Gallego}
\affiliation{ICFO-Institut de Ci\`{e}ncies Fot\`{o}niques, Av. Carl Friedrich Gauss 3, E-08860 Castelldefels (Barcelona), Spain}
\author{Ad\'an Cabello}
\affiliation{Departamento de F\'{\i}sica Aplicada II, Universidad de Sevilla, E-41012 Sevilla, Spain}
\affiliation{Department of Physics, Stockholm University, S-10691 Stockholm, Sweden}
\author{Antonio Ac\'in}
\affiliation{ICFO-Institut de Ci\`{e}ncies Fot\`{o}niques, Av. Carl Friedrich Gauss 3, E-08860 Castelldefels (Barcelona), Spain}
\affiliation{ICREA-Instituci\'o Catalana de Recerca i Estudis Avan\c cats, Lluis Companys 23, E-08010 Barcelona, Spain}

\begin{abstract}
Local measurements on bipartite maximally entangled states can yield correlations that are maximally nonlocal, monogamous, and with fully random outcomes. This makes
these states ideal for bipartite cryptographic tasks. Genuine-multipartite nonlocality constitutes a stronger notion of nonlocality in the
multipartite case. Maximal genuine-multipartite nonlocality, monogamy, and random outcomes are thus highly desired properties for
 genuine-multipartite cryptographic scenarios. We prove that local measurements on any Greenberger-Horne-Zeilinger state can produce correlations that are fully genuine-multipartite nonlocal, monogamous and with fully random outcomes. A key ingredient in our proof is a multipartite chained Bell inequality detecting genuine-multipartite nonlocality, which we introduce. Finally, we discuss applications  to device-independent secret sharing.
 \end{abstract}


\maketitle
{\it Introduction.---}Local measurements on entangled quantum
states can lead to correlations that cannot be reproduced by local
models, where correlations are due to preestablished
classical variables~\cite{Bell}. This impossibility is known as
quantum nonlocality and represents one of the fundamental
properties of quantum mechanics. More recently, quantum
nonlocality has acquired the status of a resource, due to its
application for quantum information purposes~\cite{ekert}, in
particular for quantum key distribution against eavesdroppers with nonsignaling capabilities~\cite{bhk}, device-independent
quantum key distribution~\cite{diqkd},  and randomness
generation~\cite{rgen}. 

Given some correlations between the measurement results on two
parts, the nonlocal fraction \cite{epr2} quantifies the number
of events that cannot be described by a local model. As such, it
can be taken as a measure of nonlocality: While this fraction is
zero for local correlations, {\it maximally nonlocal correlations}
are such that the nonlocal fraction is equal to one. The violation
of any Bell inequality sets a lower bound on the nonlocal fraction
of the corresponding correlations \cite{BKP}. Any correlations are
maximally nonlocal if, and only if, they attain the maximal
violation, over all nonsignaling correlations, of some Bell inequality.
Apart from maximal nonlocality, another extreme property of
correlations is that of {\it monogamy} with respect to general
nonsignaling correlations. Any given (nonsignaling) $N$-partite
correlations are monogamous if the only nonsignalling
extension of them to $N+1$ parts is the trivial one in which the
part $N+1$ is uncorrelated to the initial $N$ parts. Monogamy of
correlations is clearly a very desirable property for
cryptographic purposes. Note, however, that local deterministic
correlations are monogamous but useless for cryptography. This is
where the third ingredient comes into play: {\it randomness}. The
correlations have to be such that the local outcomes are fully
unpredictable by an adversary. A nonlocal fraction of unity is
necessary but not sufficient both for the monogamy and full
randomness of nonlocal correlations. 

In Ref.~\cite{BKP}, Barrett, Kent, and Pironio showed that bipartite maximally entangled states can yield maximally
nonlocal and monogamous correlations with fully random outcomes. They first exploited
the fact that these states maximally violate the chained
inequality~\cite{chained}, which implies that the nonlocal
fraction is one. Then, contrary to other examples of bipartite maximally
nonlocal correlations~\cite{refs}, they proved that the
correlations leading to the maximal violation of the chained
inequality also have the properties of being monogamous and having fully random local outcomes.

In a general multipartite scenario with $N$ parts, these questions have hardly been considered (see,
however,~\cite{Almeida}). The multipartite situation  is conceptually richer, as
apart from the bare division between local and nonlocal,
correlations allow for finer subclassifications in terms of locality
among the different partitions. Indeed, one can consider $k$-local
models in which the $N$ parts are split into $k<N$ groups such
that (i) the parties within each group can make use of any
nonlocal resource, but (ii) the $k$ groups are only classically 
correlated. Any  correlations that can be reproduced by
these models do not contain {\it genuine-multipartite
nonlocality}, as nonlocal resources among only subsets of the $N$
parts suffice. As in the case of locality in
the bipartite setting, it is possible to construct inequalities to
detect genuine-multipartite nonlocality, known as Svetlichny
inequalities~\cite{Svetlichny}. A maximal violation of a Svetlichny inequality
implies that the corresponding correlations are maximally
genuinely multipartite nonlocal.

It was an open question whether there exist fully genuinely
multipartite nonlocal correlations with a quantum realization~\cite{Almeida}. In
this work, we prove that this is the case: Fully genuine-multipartite nonlocal correlations can be derived from
Greenberger-Horne-Zeilinger (GHZ) states \cite{ghz} of any number $N$
of parts and local dimension $d$. To this end, we construct a family
of Svetlichny inequalities generalizing the bipartite chained
inequality, and show that GHZ states attain the algebraic
violation in the limit of an infinite number of measurements.
Then, we prove that the corresponding nonlocal quantum
correlations are monogamous and fully random in the sense that the outcomes of any choice of $m<N$ parts provides $m$ perfect random dits. Finally, we draw
some implications on device-independent secret sharing.

Before proceeding, it is worth mentioning the relation between our work and Ref.~\cite{Almeida}.
There, criteria for the detection of quantum states with
 maximally genuine-multipartite correlations were provided. Using
these criteria, it was shown that all graph states lead to such
extremal nonlocal correlations. However, there genuine-multipartite nonlocality was studied with respect to $k$-local
models in which the correlations among parties within each of the
$k$ groups are, for each value of the hidden variable, nonsignaling. In contrast, the $k$ local models
considered here are the most general ones, as no constraint is
imposed on the correlations among parties within the same group.
In this fully general scenario, no example of fully genuine-multipartite nonlocal correlations with quantum realization was
known. Moreover, monogamy and randomness in a
general multipartite scenario had not been   considered previously
either.

{\it Multipartite Svetlichny chained inequalities.---}Let us
start by deriving the Bell inequality used in the proof of our
results. The bipartite chained Bell inequality for $M$ settings
and $d$ outcomes can be expressed as \cite{BKP}
\begin{eqnarray}
\label{I2}
I_M^2\doteq\sum_{\alpha=1}^{M}\big(\langle[A_\alpha-B_{\alpha}]_d\rangle+\langle[B_{\alpha}-A_{\alpha+1}]\rangle\big)\geq d-1,
\end{eqnarray}
where $\langle \Omega_{\alpha} \rangle$ stands for the average
$\sum_{i=1}^{d-1}iP(\Omega_{\alpha}=i)$, with $P(\Omega_{\alpha}=i)$ the
probability that random variable $\Omega_{\alpha}$ is observed to take
value $i$,  $[\Omega_{\alpha}]$ is $\Omega_{\alpha}$ modulo $d$, and
$A_{M+1}\doteq [A_1+1]$. When the measured variable is clear from
the context we will sometimes use the two different notations
$P(\Omega_{\alpha}=i)$ and $P(i|\alpha)$ interchangeably for the same
probability, depending upon convenience. Inequality \eqref{I2} is
satisfied by all local correlations and algebraically violated by
the correlations of the maximally entangled state
$\ket{\Psi^{2}_d}\doteq\frac{1}{\sqrt{d}}\sum_{q=0}^{d-1}\ket{qq}$
\cite{BKP,CollbeckRenner}. More precisely, measuring the quantum
observables
$\hat{A}_{\alpha}\doteq\sum_{r_{A_{\alpha}}=0}^{d-1}r_{A_{\alpha}}\ket{r_{A_{\alpha}}}\bra{r_{A_{\alpha}}}$
and
$\hat{B}_{\beta}\doteq\sum_{r_{B_{\beta}}=0}^{d-1}r_{B_{\beta}}\ket{r_{B_{\beta}}}\bra{r_{B_{\alpha}}}$,
where
\begin{eqnarray}
\label{eigenstates}
\nonumber\ket{r_{A_{\alpha}}}&\doteq&\frac{1}{\sqrt{d}}\sum_{q=0}^{d-1}e^{\frac{2\pi
i}{d}q(r_{A_{\alpha}}-\frac{\alpha-1/2}{M})}\ket{q},\\
\text{and}\ \ \ket{r_{B_{\beta}}}&\doteq&\frac{1}{\sqrt{d}}\sum_{q=0}^{d-1}e^{-\frac{2\pi
i}{d}q(r_{B_{\beta}}-\frac{\beta}{M})}\ket{q},
\end{eqnarray}
for
$\alpha,\beta=1,\ ...,\ M$, on $\ket{\Psi^{2}_d}$, leads to a Bell
value that for large $M$ can be well approximated as
\begin{equation}
\label{limit}
I_M^2(\Psi^2_d)\approx\frac{\pi^2}{4d^2M}\sum_{i=1}^{d-1}i/\sin^2\Big(\frac{\pi
i}{d}\Big).
\end{equation} 
This value tends to 0 as $M$ grows. Since all the terms
in \eqref{I2} are by definition non-negative, this is the maximal
violation any probability distribution can render.

Let us now extend inequality \eqref{I2} to the multipartite
scenario. We first discuss the case  $N=3$ and extend the
formalism to arbitrary $N$ later. Consider then three random
variables $A_{\alpha}$, $B_{\beta}$, and $C_{\gamma}$, for
$\alpha,\beta,\gamma=1,\ ...,\ M$, each of $d$ possible outcomes
$\{0,\ ...,\ d-1\}$, measured by  Alice, Bob, and
Charlie, respectively. The inequality
\begin{eqnarray}
\label{I3}
\nonumber
I_M^3&\doteq&\sum_{\alpha,\beta=1}^{M}\big(\langle[A_{\alpha}-B_{\alpha+\beta-1}+{C_{\beta}}]\rangle\\
\nonumber
&+&\langle[B_{\alpha+\beta-1}-A_{\alpha+1}-{C_{\beta}}]\rangle\big)\\
&\geq&
M(d-1)
\end{eqnarray} 
gives a tripartite Svetlichny-like extension of
\eqref{I2}. Here we have  introduced $B_{M+\nu}\doteq
[B_{\nu}+1]$, for any $\nu=1,\ ...,\ M$. Given  that  \eqref{I2}
is a bipartite Bell inequality, the fact that the tripartite
inequality is fulfilled by all correlations local in any bipartition
can be seen with an argument similar to one of the arguments of
\cite{Bancal}. The local relabeling $B_{\alpha}\to
B_{\alpha+\beta-1}$ of Bob's bases in $I_M^2$ gives
$I_M^2(\beta)\doteq\sum_{\alpha=1}^{M}\big(\langle[A_{\alpha}-B_{\alpha+\beta-1}]\rangle+\langle[B_{\alpha+\beta-1}-A_{\alpha+1}]\rangle\big)$.
Since this simply defines a symmetry of \eqref{I2} it also gives a
Bell inequality, $I_M^2(\beta)\geq d-1$. In turn, the $\beta$-th term in
the definition of $I_M^3$  can be recast as
$I_M^2(\beta)\circ
C_{\beta}\doteq\sum_{\alpha=1}^{M}\big(\langle[A_{\alpha}-B_{\alpha+\beta-1}-C_{\beta}]\rangle+\langle[B_{\alpha+\beta-1}-A_{\alpha+1}-C_{\beta}]\rangle\big)$,
where ``$\circ\ C_{\beta}$" stands for the ``insertion of $C_{\beta}$ with the opposite sign from
$B_{\alpha+\beta-1}$." Grouping Bob and Charlie together with a
single effective variable $B_{\alpha+\beta-1}-C_{\beta}$ we see
that, for any correlations local with respect to the bipartition
$A:BC$, it must be $I_M^2(\beta)\circ C_{\beta}\geq d-1$. In
addition, since this holds for all $\beta$ and
$I_M^3\equiv\sum_{\beta=1}^{M}I_M^2(\beta)\circ C_{\beta}$, any
correlations local with respect to $A:BC$ satisfy $I_M^3\geq
M(d-1)$. The same reasoning holds of course for correlations local
with respect to $B:AC$ and an effective variable
$A_{\alpha}+C_{\beta}$. Finally, since $I_M^3$  is symmetric with
respect to the permutation of $A$ and $C$ (see Appendix \ref{Sym}),  the tripartite inequality must be satisfied by {\it all}
probability distributions with a bilocal model. That is, by all the distributions that
can be written as a convex combination of correlations with a
local model with respect to any bipartition of the three parties. For later convenience, instead of \eqref{I3}, we
consider its regularized version $\overline{I_M^3}\doteq I_M^3/M$:
\begin{eqnarray}
\label{I3aver}
\nonumber\overline{I_M^3}&\doteq&\frac{1}{M}\sum_{\alpha,\beta=1}^{M}\big(\langle[A_{\alpha}-B_{\alpha+\beta-1}+{C_{\beta}}]\rangle\\
&+&\langle[B_{\alpha+\beta-1}-A_{\alpha+1}-{C_{\beta}}]\rangle\big)\geq d-1.
\end{eqnarray}

For arbitrary $N$, the inequality generalizes by induction as
$\overline{I_M^N}\doteq\frac{1}{M}\sum_{\psi=1}^{M}\overline{I_M^{N-1}}(\psi)\circ
Z_{\psi}$, where $Z$ is the $N$-th variable. This gives
\begin{eqnarray}
\label{IN}
\nonumber
\overline{I_M^N}&\doteq&\frac{1}{M^{N-2}}\sum_{\alpha,\beta, ...,\chi,\psi=1}^{M}\big(\langle[A_{\alpha}-B_{\alpha+\beta-1}+ ... \\
\nonumber
&-&(-1)^{N-1}Y_{\chi+\psi-1}+(-1)^{N-1}Z_{\psi}]\rangle \\
\nonumber
&+&\langle[B_{\alpha+\beta-1}-A_{\alpha+1}- ...\\
\nonumber
&+&(-1)^{N-1}Y_{\chi+\psi-1}-(-1)^{N-1}Z_{\psi}]\rangle\big)\\
&\geq& d-1,
\end{eqnarray}
for $N$ random variables $A_{\alpha}$, $B_{\beta}$, $C_{\gamma}$,
..., $Y_{\psi}$, and $Z_{\zeta}$, in possession of
Alice, Bob, Charlie, ..., Yakira, and Zack, respectively. Also, for  $\Omega=A, B, C, ..., Y,$ or $Z$, we have introduced, in a  general way, $\Omega_{i\times M+\omega}\doteq
[\Omega_{\omega}+i]$, for any integer $i$ and all $\omega=1,\ ...,\ M$. In the generic case
of arbitrary $N$, the composition rule ``$\circ$" refers to ``insertion
of the new variable with the opposite sign from the
previously inserted one." With the same reasoning \cite{Bancal} as
above, from the fact that $\overline{I_M^{N-1}}\geq d-1$ is
satisfied by all $(N-1)$-partite correlations local in at least
one bipartition, it follows by construction that \eqref{IN} is
satisfied by any $N$-partite correlations local with respect to
any bipartition ``$Z$ with at least anyone else versus the rest."
Once again, by symmetry under the permutation of $Z$ with some other
part (see Appendix \ref{Sym}), one sees that \eqref{IN} is satisfied
by all $N$-partite distributions local in a bipartition.
Equation~\eqref{IN} is the Svetlichny inequality used in what follows to
prove our results. Actually, equation~\eqref{IN} encapsulates an entire family of
Svetlichny inequalities for $M$ measurements of $d$ outcomes. The
same family of inequalities is independently derived in~\cite{geneva}.

For the quantum realization, we  introduce first Charlie's 
observable
$\hat{C}_{\gamma}\doteq\sum_{r_{C_{\gamma}}=0}^{d-1}r_{C_{\gamma}}\ket{r_{C_{\gamma}}}\bra{r_{C_{\gamma}}}$,
where
\begin{equation}
\ket{r_{C_{\gamma}}}\doteq\frac{1}{\sqrt{d}}\sum_{q=0}^{d-1}e^{\frac{2\pi
i}{d}q(r_{C_{\gamma}}-\frac{\gamma-1}{M})}\ket{q},
\end{equation} for
$\gamma=1,\ ...,\ M$. In turn, for all other users we introduce
analogous observables. For instance, for Yakira and Zack we define
$\hat{Y}_{\psi}$ and $\hat{Z}_{\zeta}$ respectively with
eigenstates
\begin{eqnarray}
\label{eigenstatesyakzack}
\nonumber\ket{r_{Y_{\psi}}}&\doteq&\frac{1}{\sqrt{d}}\sum_{q=0}^{d-1}e^{-(-1)^{N-1}\frac{2\pi
i}{d}q(r_{Y_{\psi}}-\frac{\psi-1}{M})}\ket{q}\\ 
\text{and}\ \ \ket{r_{Z_{\zeta}}}&\doteq&\frac{1}{\sqrt{d}}\sum_{q=0}^{d-1}e^{(-1)^{N-1}\frac{2\pi
i}{d}q(r_{\hat{Z}_{\zeta}}-\frac{\zeta-1}{M})}\ket{q},
\end{eqnarray}
for $\psi,\zeta=1,\ ...,\ M$. In the  limit $M\to\infty$ the maximal violation of inequality~\eqref{IN}  is obtained by measuring these observables on the $N$-partite GHZ state $\ket{\Psi^{N}_d}\doteq\frac{1}{\sqrt{d}}\sum_{q=0}^{d-1}\ket{qqq\
...\ qq}$. To see this we show in Appendix \ref{I2I3Aver} that
\begin{equation}
\label{euqalbellvalues}
\overline{I_M^2}(\Psi^2_d)=\overline{I_M^3}(\Psi^3_d)= ... =\overline{I_M^N}(\Psi^N_d),
\end{equation}
 where
$\overline{I_M^3}(\Psi^3_d)$ and $\overline{I_M^N}(\Psi^N_d)$ are, respectively, the Bell
values of  \eqref{I3aver} and \eqref{IN} for the observables defined above on states $\ket{\Psi^{3}_d}$ and
$\ket{\Psi^{N}_d}$. Thus, the Bell values for all $N$ equally  tend to
zero as $M$ grows. Since inequality \eqref{IN} consists, as in the bipartite case, exclusively 
of non-negative terms,  in the limit $M\to\infty$ GHZ states attain its algebraic violation. Furthermore,
since the inequality is only violated by genuinely multipartite
nonlocal correlations, the latter implies that all GHZ
states are maximally
genuine-multipartite nonlocal.


{\it Monogamous and fully random genuinely multipartite quantum correlations.---}Any
correlations $P$   featuring $\overline{I_M^N}(P)=0$ must
necessarily satisfy 
\begin{equation}
\label{constraints}
P(r_{A}\neq [r_{B}- ...
-(-1)^{N-1}r_{Z}],r_{B}, ...,r_{Z}|\alpha,\beta,
...,\zeta)\equiv0,
\end{equation} 
for all $r_{A}$, $r_{B}$, ...,
$r_{Z}\in\{0,1, ..., d-1\}$ and $(\alpha,\beta, ...,\zeta)$ being
any  of the 2$M^{N-1}$ measurement bases appearing in
\eqref{IN}. Note that not all possible combinations of the $M$
local bases appear in the inequality. Hereafter, we study
the properties of the probability distributions $P$ corresponding
to the bases appearing in \eqref{IN}. For each such distribution, $d^{N}-d^{N-1}$ coefficients are automatically set to zero by  \eqref{constraints}. The remaining $d^{N-1}$ are univocally determined
by the marginal probability distribution corresponding to any $N-1$ parts. For instance, 
\begin{eqnarray}
\label{instance}
\nonumber&&\sum_{r_{A}}P(r_{A},r_{B}, ...,r_{Z}|\alpha,\beta, ...,\zeta)\\
\nonumber&=&P(r_{A}=
[r_{B}- ... -(-1)^{N-1}r_{Z}],r_{B},
...,r_{Z}|\alpha,\beta, ...,\zeta)\\
&=&P(r_{B},
...,r_{Z}|\beta, ...,\zeta),
\end{eqnarray} 
and equivalently for other parties
and measurement bases. When $M\to\infty$, the following theorem
fixes the value of all $(N-1)$-partite marginals and hence imposes
uniqueness. In turn, the uniqueness of $P$  implies also its
monogamy~\cite{masanes}. Moreover, the theorem proves also the
full randomness of all its  marginal distributions.

{\theorem For any $N$-partite nonsignalling distribution $P$ such
that  $\overline{I_M^N}(P)\leq \varepsilon$, with
$\varepsilon\geq0$, the  marginal distributions fulfill
\begin{equation}
P\big(\mathcal{S}(r_{A}, ...,r_{Z})|\mathcal{S}(\alpha, ...,\zeta)\big)\leq \frac{1}{d^{N-1}}+\frac{d(N-1)}{4}\varepsilon,
\end{equation}
for $(\alpha, ...,\zeta)$ any of the settings appearing in
\eqref{IN}, where $\mathcal{S}$ refers to any subset of $N-1$
parts out of all $N$.}

Note that for $P$ realized by  GHZ states $\ket{\Psi^N_d}$ and the measurements considered here, it is $\overline{I_M^N}(P)\approx\frac{\pi^2}{4d^2M}\sum_{i=1}^{d-1}i/\sin^2(\frac{\pi
i}{d})$, which tends to 0 as $M$ grows. Therefore, the GHZ-state quantum realization fulfills the theorem for any arbitrarily small $\varepsilon$. In this limit,  the  theorem thus guarantees that $P\big(\mathcal{S}(r_{A}, ...,r_{Z})|\mathcal{S}(\alpha,...,\zeta)\big)= \frac{1}{d^{N-1}}$. That is, that all the $(N-1)$-partite marginal distributions (and therefore  all the marginal distributions) have each and all of their outcomes equally probable, or in other words, that they are fully random.

{\it Device-independent secret sharing.---}Monogamy of
multipartite correlations is a desired property in multipartite
cryptographic scenarios. In particular, for instance, if
$\overline{I_M^N}(P)=0$ then  correlations $P$ fulfill the
requirements for a device-independent implementation of the
quantum secret-sharing protocol introduced in
\cite{secretsharing}.  We analyze this for the particular case
$N=3$ for ease of notation, but the same conclusions are valid for
any $N\geq3$. Alice wishes to share secret dits with Bob and
Charlie, but she suspects that one of them is dishonest.
Therefore, she wishes to do it in such a way that Bob and Charlie can access
the value of  the dits only if they are together. The
three distant users then randomly input settings $\alpha$, $\beta$,
and $\gamma$ into three black boxes described by correlations
$P(r_A,r_B,r_C|\alpha, \beta,\gamma)$  with the property that
$\overline{I_M^3}(P)=0$. They repeat the procedure many times,
each time recording the outcome, and at the end publicly broadcast
all the settings used. From Theorem 1 they know that whenever
their settings happen to match those of \eqref{I3aver},  i.e., 
$\alpha-\beta+\gamma-1=0$ (modulo $M$), or $\alpha-\beta+\gamma=0$
(modulo $M$), then $P(r_{A}=a| [r_{C}-r_{B}]=a)\equiv1$ for all
$a\in\{0,1, ..., d-1\}$. This means that then, if Bob and Charlie
meet, they can determine with certainty the value of Alice's dit
$r_A$ simply by subtracting their outcomes. In addition, since all
marginals of $P$ (for the relevant settings) are fully random, neither
Bob nor Charlie can obtain any information at all from their local
outcomes alone. Finally, as the correlations are monogamous,
Alice's dit is also unpredictable by any external adversary.


{\it Conclusions.---}We presented a multipartite version of the
multiple-setting multiple-outcome chained Bell inequalities. The
inequalities introduced are Svetlichny-like: they are
satisfied by all probability distributions expressed as
mixtures of local correlations with respect to any bipartition. We
showed that, in the limit of an infinite number of settings,
 correlations from GHZ states of any local dimensions or numbers
of parts violate these inequalities as much as any 
nonsignaling correlations. This proves that  the
genuine-multipartite nonlocal content of GHZ states is maximal.
Moreover, we showed that any correlations algebraically violating
the present inequalities are monogamous with respect to nonsignaling compositions 
and yield fully random outcomes for any subset of parts. This proves monogamy and full randomness of genuinely multipartite quantum correlations in a nonsignaling scenario. 
Finally, we showed that the correlations from GHZ
states approach,  as the number of measurement settings grows,
those required for device-independent secret sharing secure
against eavesdroppers limited solely by the no-signalling
principle.


{\bf Sketch of the proof of Theorem 1}. To end up with, we provide here just the main steps of the proof, the most technical calculations being detailed
in the appendices.

The proof for arbitrary $N\geq2$ is in a similar spirit to the proof given in \cite{BKP} for the particular case $N=2$. We proceed by {\it reductio ad
absurdum}. We start by the particular marginal $P(r_{A}, ...
,r_{Y}|\alpha,\alpha+\beta-1, ..., \chi+\psi-1)$, corresponding to
all parts but $Z$, and assume that for some input
$(\alpha',\alpha'+\beta'-1, ..., \chi'+\psi'-1)$ the most probable
outcome $(a^{max}_{(\alpha', ..., \psi')}, ..., y^{max}_{(\alpha',
..., \psi')})$ is such that 
\begin{eqnarray}
\label{assumption}
\nonumber
&&\!\!\!\!\!\!\!\!P(a^{max}_{(\alpha', ..., \psi')},
..., y^{max}_{(\alpha', ..., \psi')}|\alpha',\alpha'+\beta'-1,
..., \chi'+\psi'-1)\\
&&\!\!\!\!\!\!\!\!>1/d^{N-1}+\frac{d(N-1)}{4}\varepsilon.
\end{eqnarray} 
Then, we prove that this implies that $\overline{I_M^N}(P)>
\varepsilon$, which contradicts the  hypothesis. [The same assumption for
$(\alpha'+1,\alpha'+\beta'-1, ..., \chi'+\psi'-1)$ would
lead to an equivalent contradiction.] Finally, we extend the proof to the other $(N-1)$-partite marginals by symmetry.

First, since
\begin{equation}
\label{W}
\langle[W]\rangle\doteq\sum_{i=1}^{d-1}iP\big([W]=i)\geq1-P\big([W]=0),
\end{equation}
we see from \eqref{IN} that
\begin{eqnarray}
\nonumber 
&&\overline{I_M^N}\geq\\
\nonumber
&&2M-\frac{1}{M^{N-2}}\sum_{\alpha,\beta,
...,\chi,\psi=1}^{M}\Big(P\big(A_{\alpha}=[B_{\alpha+\beta-1}- ... +
\\
\nonumber&&(-1)^{N-1}Y_{\chi+\psi-1}-(-1)^{N-1}Z_{\psi}]\big)+
P\big(A_{\alpha+1}=\\
\nonumber&&[B_{\alpha+\beta-1}- ...
+(-1)^{N-1}Y_{\chi+\psi-1}-(-1)^{N-1}Z_{\psi}]\big)\Big).\\
&&
\end{eqnarray}

Next, we notice in Appendix \ref{Condassum} that, for all $(\alpha,
...,\omega)$,  
\begin{eqnarray}
\label{Condassum0}
\nonumber 
&&P\big(A_{\alpha}=[B_{\beta}- ...
+(-1)^{N-1}Y_{\psi}-(-1)^{N-1}Z_{\zeta}]\big)\leq\\
\nonumber 
&&1-\big|P(A_{\alpha}=a,B_{\beta}=b,
... ,Y_{\psi}=y)-P\big(B_{\beta}=b, ... ,Y_{\psi}\\
&&=y,Z_{\zeta}=[y-
... +(-1)^{N-1}b-(-1)^{N-1}a]\big)\big|,
\end{eqnarray} for any $(a, b, ...,y)$.
Then,
\begin{eqnarray}
\nonumber &&\overline{I_M^N}\\
\label{caca}
\nonumber&\geq&\frac{1}{M^{N-2}}\times\\
\nonumber&&\sum_{\alpha,\beta, ...,\chi,\psi=1}^{M}\big|P\big(A_{\alpha}=a,B_{\alpha+\beta-1}=b, ...,Y_{\chi+\psi-1}=y\big)\\
&-&P\big(A_{\alpha+1}=a,B_{\alpha+\beta-1}=b, ...,Y_{\chi+\psi-1}=y\big)\big|,
\end{eqnarray}
where the triangle inequality has been used.

In Appendix \ref{CondIstar}, in turn, we see that  hypothesis \eqref{assumption} implies that
there exists some point $({a_0}_{(\alpha', ..., \psi')}, ...,
{y_0}_{(\alpha', ..., \psi')})$ in the $d^{N-1}$-dimensional cubic grid $\mathcal{G}\doteq\{0,1, ..., d-1\}^{\times (N-1)}$, such that
\begin{eqnarray}
\label{Istar}
\nonumber\big|P(A_{\alpha'}={a_0}_{(\alpha', ..., \psi')}, ...,Y_{\chi'+\psi'-1}= {y_0}_{(\alpha', ..., \psi')})\!\!&-&\\
P(A_{\alpha'}=\dot{a_0}_{(\alpha', ..., \psi')}, ...,Y_{\chi'+\psi'-1}=\dot{{y_0}}_{(\alpha', ...,\psi')})\big|\!\!&>&\!\varepsilon,\ \ \ \ \ \
\end{eqnarray}
where $(\dot{a_0}_{(\alpha', ..., \psi')}, ...,
\dot{{y_0}}_{(\alpha', ..., \psi')})\ \in\mathcal{G}$ is any nearest neighbor of
$({a_0}_{(\alpha', ..., \psi')}, ..., {y_0}_{(\alpha', ...,
\psi')})$.

Now,  defining $\tilde{\beta}\doteq \alpha+\beta$, ...,
$\tilde{\chi}\doteq \varphi+\chi$  and $\tilde{\psi}\doteq
\chi+\psi$ in \eqref{caca}, we see that
\begin{eqnarray}
\label{noname}
\nonumber&&\overline{I_M^N}\geq\frac{1}{M^{N-2}}\sum_{\tilde{\psi},...
,\tilde{\beta}=1}^{M}\\
\nonumber&&\Big|\sum_{\alpha=1}^{M}\big(P(A_{\alpha}=a,B_{\tilde{\beta}-1}=b,...\,Y_{\tilde{\psi}-1}=y)\\
&-&P(A_{\alpha+1}=a,B_{\tilde{\beta}-1}=b, ...,Y_{\tilde{\psi}-1}=y)\big)\Big|.
\end{eqnarray}
Here,  we choose $b\equiv {b_0}_{(\alpha', ..., \psi')}$, ..., and
$y\equiv {y_0}_{(\alpha', ..., \psi')}$. In turn,  we set
$a={a_0}_{(\alpha', ..., \psi')}$, for all
$1\leq\alpha\leq\alpha'$, and $a={a_0}_{(\alpha', ..., \psi')}+1$,
for all $\alpha'+1\leq\alpha\leq M$. With this, inequality \eqref{noname} becomes 
\begin{eqnarray}
\label{noname2}
\nonumber\overline{I_M^N}
&\geq&\frac{1}{M^{N-2}}\sum_{\tilde{\psi},...
,\tilde{\beta}=1}^{M}\\
\nonumber&&\big|(P(A_{\alpha}={a_0}_{(\alpha', ...,
\psi')},B_{\tilde{\beta}-1}={b_0}_{(\alpha', ..., \psi')}, ...,\\
\nonumber
&&Y_{\tilde{\psi}-1}={y_0}_{(\alpha', ...,
\psi')})-P(A_{\alpha}={a_0}_{(\alpha', ...,
\psi')}+1,\\
\nonumber
&&B_{\tilde{\beta}-1}={b_0}_{(\alpha', ..., \psi')}, ...,
Y_{\tilde{\psi}-1}={y_0}_{(\alpha', ..., \psi')})\big|\\
\nonumber
&>&
\frac{1}{M^{N-2}}\sum_{\tilde{\psi},...
,\tilde{\beta}=1}^{M}\varepsilon\\
&=&\varepsilon,
\end{eqnarray}
 where we have used that
$A_{M+1}=[A_{1}+1]$ and invoked property \eqref{Istar}. The last inequality
finishes the proof for marginal $P(r_{A}, ...
,r_{Y}|\alpha,\alpha+\beta-1, ..., \chi+\psi-1)$. 

The proof for
any other $(N-1)$-party marginal that includes  $A$ is a replica
but where, before  \eqref{caca}, instead of
grouping it together with $B$, $C$, ..., and $Y$, one groups $A$
with any choice of $N-2$ out of the other $N-1$. Finally, the
proof for the marginal $P\big(r_{B}, ...,
r_{Z}|\alpha,\alpha+\beta-1, ..., \chi+\psi-1)$, follows due to
invariance of $\overline{I_M^N}$ under the exchange of $A$ with, for instance, $C$ (see Appendix \ref{Sym}). $\square$

{\it Acknowledgments.---}We acknowledge support from  
the Spanish projects FIS2008-05596, FIS2010-14830, FIS2011-29400, and QOIT (Consolider Ingenio 2010), Chist-Era DIQIP, the European
 EU FP7 Projects Q-Essence
and QCS, and an ERC Starting Grant PERCENT, the Juan de la Cierva
Foundation, Catalunya Caixa, Generalitat de Catalunya, and the
Wenner-Gren Foundation.



\appendix

\section{Symmetry under permutations of parts}
\label{Sym}
Here we show the invariance of the $N$-partite inequality \eqref{IN} under certain  permutations of parts, the same arguments holding also for the  tripartite case of \eqref{I3aver}.
 We show explicitly that \eqref{IN} is symmetric with respect to the exchange of the $N$-th and the $(N-2)$-th parts, $Z$ and $X$. The proof for the exchange $A\leftrightarrow C$ is exactly 
the same. We write the Bell polynomial of  \eqref{IN} as
\begin{widetext}
\begin{eqnarray}
&&\overline{I_M^N}\doteq\frac{1}{M^{N-2}}\sum_{\alpha,\beta, ...,\varphi,\chi, \psi=1}^{M}\big(J_{\alpha, ...,\varphi,\chi, \psi}(A, B, ..., X, Y, Z)+H_{\alpha, ...,\varphi,\chi, \psi}(A, B, ..., X, Y, Z)\big),\\
\nonumber
&&\text{with }J_{\alpha, ...,\varphi,\chi, \psi}(A, B, ..., X, Y, Z)\doteq\langle[A_{\alpha}-B_{\alpha+\beta-1}\ ... +(-1\big)^{N-1}X_{\varphi+\chi-1}-(-1\big)^{N-1}Y_{\chi+\psi-1}+(-1\big)^{N-1}Z_{\psi}]\rangle,\\
\nonumber
&&\text{and } \ H_{\alpha, ...,\varphi,\chi, \psi}(A, B, ..., X, Y, Z)\doteq\langle[B_{\alpha+\beta-1}-A_{\alpha+1}\ ... +(-1\big)^{N}X_{\varphi+\chi-1}-(-1\big)^{N}Y_{\chi+\psi-1}+(-1\big)^{N}Z_{\psi}]\rangle.
\end{eqnarray}
\end{widetext}
Under the exchange  $X\leftrightarrow Z$, these matrices transform as 
\begin{widetext}
\begin{eqnarray}
\nonumber&&J_{\alpha, ...,\varphi,\chi, \psi}(A, B, ..., X, Y, Z)\rightarrow J_{\alpha, ..., \varphi,\chi, \psi}(A, B, ..., Z, Y, X)\equiv J_{\alpha, ...,\varphi, \psi-\varphi+1, \varphi+\chi-1}(A, B, ...,X, Y, Z)\ \text{and}\\ 
 &&H_{\alpha, ..., \varphi, \chi, \psi}(A, B, ..., X, Y, Z)\rightarrow H_{\alpha, ..., \varphi, \chi, \psi}(A,B, ...,Z, Y,X)\equiv H_{\alpha, ..., \varphi, \psi-\varphi+1, \varphi+\chi-1}(A, B, ..., X, Y, Z). 
\end{eqnarray}
\end{widetext}
We notice that, due to the symmetries in the definition of matrix $J$, the fact that $\Omega_{i\times M+\omega}\doteq
[\Omega_{\omega}+i]$, for any $\Omega=A, B, C, ..., Y,$ or $Z$, implies that $J_{\alpha, ...,\omega\pm M, ...,\varphi,\chi, \psi}(A, B, ..., X, Y, Z)=J_{\alpha, ...,\omega, ...,\varphi,\chi,\psi}(A,B, ..., X, Y, Z)$, for any 
$\omega=\alpha, \beta, ...\varphi, \chi$ or $\psi$. Analogously, the same property holds also for matrix $H$. Hence, we have that 
\begin{eqnarray}
\nonumber&&\!\!\!\!\!\!\!\!\!\!\!\!\!\!\!\!\!\!\!\!\sum_{\alpha,\beta, ..., \varphi, \chi, \psi=1}^{M}J_{\alpha, ..., \varphi, \chi, \psi}(A, B, ..., X, Y, Z)\equiv\\
&&\!\!\!\!\!\!\!\!\!\!\!\!\!\!\!\!\!\!\!\!\sum_{\alpha, \beta, ..., \varphi, \chi, \psi=1}^{M}J_{\alpha, ..., \varphi, \psi-\varphi+1, \varphi+\chi-1}(A, B, ..., X, Y, Z)
\end{eqnarray}
and
\begin{eqnarray}
\nonumber&&\!\!\!\!\!\!\!\!\!\!\!\!\!\!\!\!\!\!\!\!\sum_{\alpha,\beta, ...,\varphi,\chi, \psi=1}^{M}H_{\alpha, ...,\varphi,\chi, \psi}(A, B, ..., X, Y, Z)\equiv\\
&&\!\!\!\!\!\!\!\!\!\!\!\!\!\!\!\!\!\!\!\!\sum_{\alpha,\beta, ...,\varphi,\chi, \psi=1}^{M}H_{\alpha\ ...,\varphi, \psi-\varphi+1, \varphi+\chi-1}(A, B, ..., X, Y, Z).
\end{eqnarray}
Therefore, it is $\overline{I_M^N}\equiv\overline{I_M^N}(X\leftrightarrow Z)$. $\square$
\section{Equality \eqref{euqalbellvalues}}
\label{I2I3Aver}
Here we prove that, for any $N>2$, it is $\overline{I_M^N}(\Psi^N_d)=\overline{I_M^{N-1}}(\Psi^{N-1}_d)$. Consider first the expectation value 
\begin{eqnarray}
\nonumber&&\!\!\!\!\bra{\Psi^{N-1}_d}[\hat{A}_{\alpha}-\hat{B}_{\alpha+\beta-1}+ ... -(-1)^{N-1}\hat{Y}_{\chi}]\ket{\Psi^{N-1}_d}\equiv\\
\nonumber&&\!\!\!\!\!\!\!\!\!\!\!\!\sum_{r_{A_{\alpha}},r_{B_{\alpha+\beta-1}}, ..., r_{Y_{\chi}}=0}^{d-1}[r_{A_{\alpha}}-r_{B_{\alpha+\beta-1}}+ ... -(-1)^{N-1}r_{Y_{\chi}}]\\
\nonumber&&\!\!\!\!\big|\bra{r_{A_{\alpha}}}\bra{r_{B_{\alpha+\beta-1}}} ... \bra{r_{Y_{\chi}}}\Psi^{N-1}_d\rangle\big|^2\\
&=&\frac{1}{d^{N}}\times d^{N-2}\sum_{n=1}^{d-1}n\Big|\frac{1-e^{-\pi i/M}}{1-e^{-\frac{2\pi i}{d}(n+1/2)}}\Big|^2,
\end{eqnarray}
where we have used the explicit definitions of $\ket{r_{A_{\alpha}}}$, $\ket{r_{B_{\alpha+\beta-1}}}$, ..., and $\ket{r_{Y_{\chi}}}$,  summed a geometric sequence, and introduced $n\equiv r_{A_{\alpha}}-r_{B_{\alpha+\beta-1}}+ ... -(-1\big)^{N-1}r_{Y_{\chi}}$. Consider next 
\begin{eqnarray}
\nonumber&&\!\!\!\!\bra{\Psi^{N}_d}[\hat{A}_{\alpha}-\hat{B}_{\alpha+\beta-1}+ ... +(-1)^{N-1}\hat{Z}_{\psi}]\ket{\Psi^{N}_d}=\\
\nonumber&&\!\!\!\!\!\!\!\!\!\!\!\!\sum_{r_{A_{\alpha}},r_{B_{\alpha+\beta-1}}, ..., r_{Z_{\psi}} =0}^{d-1}[r_{A_{\alpha}}-r_{B_{\alpha+\beta-1}}+ ... +(-1)^{N-1}r_{Z_{\psi}}]\\\nonumber&&\!\!\!\!\big|\bra{r_{A_{\alpha}}}\bra{r_{B_{\alpha+\beta-1}}} ... \bra{r_{Z_{\psi}}}\Psi^{N}_d\rangle\big|^2\\
&=&\frac{1}{d^{N+1}}\times d^{N-1}\sum_{n'=1}^{d-1}n'\Big|\frac{1-e^{-\pi i/M}}{1-e^{-\frac{2\pi i}{d}(n'+1/2)}}\Big|^2,
\end{eqnarray}
where we have now also used the definition of $\ket{r_{Z_{\psi}}}$, and introduced $n'\equiv r_{A_{\alpha}}-r_{B_{\alpha+\beta-1}}+ ... +(-1\big)^{N-1}r_{Z_{\psi}}$. Notice that both expectation values coincide for any $\alpha$, $\beta$, ..., $\chi$, and $\psi$. In addition, the same analysis holds true for $\bra{\Psi^{N-1}_d}[\hat{B}_{\alpha+\beta-1}-\hat{A}_{\alpha+1}- ... +(-1\big)^{N-1}\hat{Y}_{\chi}]\ket{\Psi^{N-1}_d}$  and $\bra{\Psi^{N}_d}[\hat{B}_{\alpha+\beta-1}-\hat{A}_{\alpha+1}- ... -(-1)^{N-1}\hat{Z}_{\psi}]\ket{\Psi^{N}_d}$. Therefore, the Bell value of the $\psi$-th term of $\overline{I^N_M}$, $\frac{1}{M^{N-2}}\sum_{\alpha,\beta, ...,\chi=1}^{M}\big(\langle[A_{\alpha}-B_{\alpha+\beta-1}+ ... -(-1)^{N-1}Y_{\chi+\psi-1}+(-1)^{N-1}Z_{\psi}]\rangle+\langle[B_{\alpha+\beta-1}-A_{\alpha+1}- ... +(-1)^{N-1}Y_{\chi+\psi-1}-(-1)^{N-1}Z_{\psi}]\rangle\big)$, obtained from quantum measurements on $\ket{\Psi^{N}_d}$, is equal to $1/M$ times the Bell value of $\overline{I^{N-1}_M}$  obtained from $\ket{\Psi^{N-1}_d}$. Summing over $\psi$ completes the proof. $\square$
\section{Bound \eqref{Condassum0}}
\label{Condassum}
Here we  show that, for all $(\alpha,\beta, ..., \psi, \zeta)$, and any $(a,b, ..., y)$, bound \eqref{Condassum0} holds. One has
\begin{widetext}
\begin{eqnarray}
\nonumber
& &P\big(A_{\alpha}=B_{\beta}- ... +(-1)^{N-1}Y_{\psi}-(-1)^{N-1}Z_{\zeta})\\
\nonumber
&\doteq&\sum_{i,j, ..., m}P\big(A_{\alpha}=i,B_{\beta}=j, ... ,Y_{\psi}=m,Z_{\zeta}=m- ... +(-1)^{N-1}j-(-1)^{N-1}i\big)\\
 \nonumber
&\leq&\sum_{i,j, ..., m}\min\Big(P(A_{\alpha}=i,B_{\beta}=j, ... ,Y_{\psi}=m),P\big(B_{\beta}=j, ... ,Y_{\psi}=m,Z_{\zeta}=m- ... +(-1)^{N-1}j-(-1)^{N-1}i\Big)\\
 \nonumber
&\leq&\min\Big(P(A_{\alpha}=a,B_{\beta}=b, ... ,Y_{\psi}=y),P\big(B_{\beta}=b, ... ,Y_{\psi}=y,Z_{\zeta}=y- ... +(-1)^{N-1}b-(-1)^{N-1}a\Big)\\
\nonumber
&+&\min\Big(\sum_{(i,j, ..., m)\neq (a,b, ... ,y)}P(A_{\alpha}=a,B_{\beta}=b, ... ,Y_{\psi}=y),\\
\nonumber
& &\sum_{(i,j, ..., m)\neq (a,b, ... ,y)}P\big(B_{\beta}=b, ... ,Y_{\psi}=y,Z_{\zeta}=y- ... +(-1)^{N-1}b-(-1)^{N-1}a\big)\Big)\\
 \nonumber
 &\equiv&\min\Big(P\big(A_{\alpha}=a,B_{\beta}=b, ... ,Y_{\psi}=y),P\big(B_{\beta}=b, ... ,Y_{\psi}=y,Z_{\zeta}=y- ... +(-1)^{N-1}q-(-1)^{N-1}a\big)\Big)\\
\nonumber
&+&\min\Big(1-P(A_{\alpha}=a,B_{\beta}=b, ... ,Y_{\psi}=y),1-P\big(B_{\beta}=b, ... ,Y_{\psi}=y,Z_{\zeta}=y- ... +(-1)^{N-1}b-(-1)^{N-1}a\big)\big)\\
 &\equiv&1-\big|P(A_{\alpha}=a,B_{\beta}=b, ... ,Y_{\psi}=y)-P\big(B_{\beta}=b, ... ,Y_{\psi}=y,Z_{\zeta}=y- ... +(-1)^{N-1}b-(-1)^{N-1}a\big)\big|,
 \end{eqnarray}
 \end{widetext}
for  arbitrary $a$, $b$, ..., $y$ $\in\ \{0,\ ...\ ,d-1\}$. $\square$
\section{Condition \eqref{Istar}}
\label{CondIstar}

Here we prove that, if for some setting $(\alpha',\alpha'+\beta'-1, ..., \chi'+\psi'-1)$ the highest probability
 $P(a^{max}_{(\alpha', ..., \psi')}, ..., y^{max}_{(\alpha', ..., \psi')}|\alpha',\alpha'+\beta'-1, ..., \chi'+\psi'-1)$ is bounded from below as in \eqref{assumption}, then
inequality \eqref{Istar} is true. Again, we proceed by {\it reductio ad absurdum}: Suppose \eqref{Istar} is false. Then,
\begin{eqnarray}
\nonumber
\big|P(A_{\alpha'}={a_0}_{(\alpha', ..., \psi')}, ...,Y_{\chi'+\psi'-1}= {y_0}_{(\alpha', ..., \psi')})\!\!&-&\\
P(A_{\alpha'}=\dot{a_0}_{(\alpha', ..., \psi')}, ...,Y_{\chi'+\psi'-1}=\dot{{y_0}}_{(\alpha', ..., \psi')})\big|\!\!&\leq&\!\varepsilon,\ \ \ \ \ \
\end{eqnarray}
for all points $({a_0}_{(\alpha', ..., \psi')}, ..., {y_0}_{(\alpha', ..., \psi')})\ \in\mathcal{G}$, where $(\dot{a_0}_{(\alpha', ..., \psi')}, ..., \dot{y_0}_{(\alpha', ..., \psi')})$ is any other point of $\mathcal{G}$ whose distance $\mathcal{D}({a_0}_{(\alpha', ..., \psi')}, ..., {y_0}_{(\alpha', ..., \psi')};\dot{a_0}_{(\alpha', ..., \psi')}, ..., \dot{y_0}_{(\alpha', ..., \psi')})$ from $({a_0}_{(\alpha', ..., \psi')}, ..., {y_0}_{(\alpha', ..., \psi')})$ is one: 
\begin{widetext}
\begin{eqnarray}
\nonumber
&&\mathcal{D}({a_0}_{(\alpha', ..., \psi')}, ..., {y_0}_{(\alpha', ..., \psi')};\dot{a_0}_{(\alpha', ..., \psi')}, ..., \dot{y_0}_{(\alpha', ..., \psi')})\equiv\mathcal{D}({a_0}_{(\alpha', ..., \psi')};\dot{a_0}_{(\alpha', ..., \psi')})+ ... +\mathcal{D}({y_0}_{(\alpha', ..., \psi')};\dot{y_0}_{(\alpha', ..., \psi')})\\
&\!\!\!\doteq& |{a_0}_{(\alpha', ..., \psi')}-\dot{a_0}_{(\alpha', ..., \psi')}|+ ... + |{y_0}_{(\alpha', ..., \psi')}-\dot{y_0}_{(\alpha', ..., \psi')}|=1.
\end{eqnarray}
\end{widetext}
This, in turn, implies that
\begin{widetext}
\begin{eqnarray}
\nonumber
&&\sum_{a,... ,y=0}^{d}P(A_{\alpha'}=a, ... , Y_{\chi'+\psi'-1}=y)\geq \sum_{a,... ,y=0}^{d}\big(P(A_{\alpha'}=a^{max}_{(\alpha', ..., \psi')}, ... , Y_{\chi'+\psi'-1}=y^{max}_{(\alpha', ..., \psi')})\\
\nonumber
&-&\varepsilon \mathcal{D}(a, ..., y;a^{max}_{(\alpha', ..., \psi')}, ..., y^{max}_{(\alpha', ..., \psi')})\big)>1+\frac{d^{N}(N-1)}{4} \varepsilon-\varepsilon(N-1)d^{N-2}\sum_{i=D^{-}_{(d)}}^{D^{+}_{(d)}}\mathcal{D}(i;0)\\
\label{sums}
&=&1+d^{N-2}(N-1)\varepsilon\bigg(\frac{d^2}{4}-\sum_{i=D^{-}_{(d)}}^{D^{+}_{(d)}}\mathcal{D}(i;0)\bigg),
\end{eqnarray}
\end{widetext}
where we have introduced $D^{+}_{(d)}\doteq (d-1)/2\doteq-D^{-}_{(d)}$, for $d$ odd, and $D^{+}_{(d)}\doteq d/2\doteq-D^{-}_{(d)}+1$, for $d$ even, and have used 
that $\mathcal{D}(a, ..., y;a^{max}_{(\alpha', ..., \psi')}, ..., y^{max}_{(\alpha', ..., \psi')})\equiv \mathcal{D}(a;a^{max}_{(\alpha', ..., \psi')})+ ... + \mathcal{D}(y;y^{max}_{(\alpha', ..., \psi')})$. Notice that, for $d$ odd, it is $\sum_{i=D^{-}_{(d)}}^{D^{+}_{(d)}}D(i;0)=(d^2-1)/4$, whereas for $d$ even it is $\sum_{i=D^{-}_{(d)}}^{D^{+}_{(d)}}D(i;0)=d^2/4$. Thus, in both cases the last line of \eqref{sums} is strictly greater than 1, which contradicts probability normalization. $\square$


\begin{thebibliography}{20}


\bibitem{Bell}
J. S. Bell, Physics \textbf{1}, 195 (1964).

\bibitem{ekert}
A. K. Ekert, Phys. Rev. Lett. \textbf{67}, 661 (1991). 

\bibitem{bhk}
J. Barrett, L. Hardy and A. Kent, Phys. Rev. Lett. \textbf{95}, 010503 (2005).

\bibitem{diqkd}
A. Ac\'in {\it et al.}, Phys. Rev. Lett. \textbf{98}, 230501 (2007).

\bibitem{rgen}
S. Pironio {\it et al.}, Nature (London) \textbf{464}, 1021 (2010); R. Colbeck, Ph.D. Thesis, University of Cambridge (2006); R. Colbeck and A. Kent, J. Phys. A: Math. Th. \textbf{44}, 095305 (2011).

\bibitem{epr2}
A. Elitzur, S. Popescu, and D. Rohrlich, Phys. Lett. A {\bf 162}, 25  (1992).

\bibitem{BKP}
J. Barret, A. Kent, and S. Pironio, Phys. Rev. Lett. {\bf 97}, 170409 (2006).

\bibitem{chained}
S. L. Braunstein and C. M. Caves,  Ann. Phys. {\bf 202}, 22 (1990).

\bibitem{refs}
 P. Heywood and M. L. G. Redhead, Found. Phys. {\bf 13}, 481 (1983);
A. Stairs, Phil. Sci. {\bf 50}, 578 (1983); A. Cabello, Phys. Rev. Lett. \textbf{87}, 010403 (2001); L. Aolita {\it et al.}, 
 arXiv: 1105.3598.

\bibitem{Almeida}
M. L. Almeida {\it et al.}, 
Phys. Rev. A {\bf 81}, 052111 (2010).

\bibitem{Svetlichny}
G. Svetlichny, Phys. Rev. D {\bf 35}, 3066 (1987); D. Collins {\it et al.}, 
Phys. Rev. Lett. {\bf 88}, 170405 (2002); M. Seevinck and G. Svetlichny, Phys. Rev. Lett. {\bf 89}, 060401 (2002).

\bibitem{ghz}
D. M. Greenberger, M. A.  Horne, and A. Zeilinger, in {\it Bell's Theorem, Quantum Theory, and Conceptions
of the Universe} (Kluwer, Dordrecht, 1989).

\bibitem{CollbeckRenner}
R. Colbeck and R. Renner,  Phys. Rev. Lett. {\bf 101}, 050403 (2008).


\bibitem{Bancal}
J.-D. Bancal {\it et al.}, 
Phys. Rev. Lett. {\bf 106}, 020405 (2011).

\bibitem{geneva}
J.-D. Bancal {\it et al.}, arXiv:1201.2055.

\bibitem{masanes}
Ll. Masanes, A. Ac\'in, and N. Gisin,
Phys. Rev. A {\bf 73,} 012112 (2006).

\bibitem{secretsharing}
M. Hillery, V. Bu{\v{z}}ek, and A. Berthiaume, Phys. Rev. A. {\bf 59}, 1829 (1999).



\end{thebibliography}
\end{document}